\documentclass[aps,prd,twocolumn,preprintnumbers,showpacs,superscriptaddress,nofootinbib,floatfix,10pt]{revtex4-1}
\pdfoutput=1
\usepackage{textcomp}
\usepackage[english]{babel}
\usepackage{amsmath,amssymb,amsbsy,booktabs}
\usepackage{bm}
\usepackage{color}
\usepackage{array}
\usepackage{amstext}
\usepackage{graphicx}
\usepackage{amsfonts}
\usepackage{dcolumn}
\usepackage{rotating}
\usepackage{epstopdf}
\usepackage{esint}
\usepackage{url}
\usepackage[colorlinks=true,
linkcolor=blue,
filecolor=blue,
anchorcolor=blue,
urlcolor=blue,
citecolor=blue
]{hyperref}

\usepackage{times}

\usepackage[utf8]{inputenc}
\usepackage{float}
\usepackage{subfigure}

\usepackage[T1]{fontenc} 

\allowdisplaybreaks

\begin{document}
\title {\Large \bf Effective neutrino number shift from keV-vacuum neutrinophilic 2HDM}

\author{Shao-Ping Li}
\email{ShowpingLee@mails.ccnu.edu.cn}
\affiliation{Institute of Particle Physics and Key Laboratory of Quark and Lepton Physics~(MOE),\\
	Central China Normal University, Wuhan, Hubei 430079, China}

\author{Xin-Qiang Li}
\email{xqli@mail.ccnu.edu.cn}
\affiliation{Institute of Particle Physics and Key Laboratory of Quark and Lepton Physics~(MOE),\\
	Central China Normal University, Wuhan, Hubei 430079, China}

\author{Xin-Shuai Yan}
\email{xinshuai@mail.ccnu.edu.cn}
\affiliation{Institute of Particle Physics and Key Laboratory of Quark and Lepton Physics~(MOE),\\
	Central China Normal University, Wuhan, Hubei 430079, China}

\author{Ya-Dong Yang}
\email{yangyd@mail.ccnu.edu.cn}
\affiliation{Institute of Particle Physics and Key Laboratory of Quark and Lepton Physics~(MOE),\\
	Central China Normal University, Wuhan, Hubei 430079, China}
\affiliation{School of Physics and Microelectronics, Zhengzhou University, Zhengzhou, Henan 450001, China}

\begin{abstract}
	If the Dirac neutrino masses are generated by a new scalar doublet with a vacuum at keV order, there would be rare hope to probe the framework in low-energy flavor physics, such as the lepton-flavor violating processes. However, the predicted neutrino Yukawa couplings being around the order of electronic Yukawa can realize a purely thermal Dirac leptogenesis, and thus render a more direct link between the high- and the low-energy CP violation. Despite its inert property in the low-energy flavor processes, 
	the keV-vacuum induced Dirac neutrino model can generate a significant shift of the effective neutrino number, which can in turn be probed by the big-bang nucleosynthesis and cosmic microwave background epochs. We show that such a keV-vacuum induced Dirac neutrino model is already constrained by the current observations and will be probed with forecast sensitivity, serving therefore as a complementary avenue to test the Dirac leptogenesis.
\end{abstract}

\pacs{}
\maketitle

\section{Introduction}

The observations of neutrino oscillations have thus far prompted several puzzles about neutrinos in the Standard Model (SM), including their tiny mass origin, their Dirac or Majorana nature, their distinctive mixing pattern from the quark sector, etc. Any trial towards these problems has catalyzed a host of investigations from theoretical constructions to experimental searches, and from low-energy particle physics to high-temperature early Universe. Despite the leading interests in Majorana neutrinos, the Dirac neutrino scenarios are also welcome and can generate promising experimental signals. If neutrinos are of the Dirac type, they can trigger the Dirac leptogenesis~\cite{Dick:1999je,Murayama:2002je} to explain the observed baryon asymmetry of the Universe (BAU). Recently, we have demonstrated that a purely thermal Dirac leptogenesis (PTDL) mechanism~\cite{Li:2020ner,Li:2021tlv} can establish a more direct link between the low-energy leptonic CP violation and the high-scale lepton asymmetry, where the BAU is formulated by the neutrino mixing without invoking specific Yukawa textures. In addition to the intimate relation with the BAU mystery, the right-handed Dirac neutrinos can also have a significant impact on the evolution of the early Universe, e.g., via generating the effective neutrino number shift, $\Delta N_{eff}$, which is constrained by the big-band nucleosynthesis (BBN)~\cite{Steigman:2012ve,Cyburt:2015mya,Pitrou:2018cgg,Fields:2019pfx} and cosmic microwave background (CMB) observations~\cite{Planck:2018vyg,Abazajian:2019oqj,Luo:2020sho,Adshead:2020ekg,Luo:2020fdt}. 

A simple and testable scenario for Dirac neutrino mass generation is by introducing a new Higgs doublet with a much smaller vacuum than the electroweak one, which is nowadays called neutrinophilic two-Higgs-doublet model (2HDM)~\cite{Gabriel:2006ns,Davidson:2009ha} (for Majorana neutrinos based on the 2HDM, see, e.g., Refs.~\cite{Atwood:2005bf,Clarke:2015hta,Li:2018rax,Li:2019xmi,Cogollo:2019mbd}). In previous investigations~\cite{Davidson:2010sf,Machado:2015sha,Bertuzzo:2015ada}, an eV-scale Higgs vacuum was widely considered to embrace $\mathcal{O}(1)$ neutrino Yukawa couplings. Phenomenological analyses of such an eV-vacuum Dirac neutrino model have also been performed therein, pointing out especially that the lepton-flavor violating (LFV) transitions, such as  $\mu\to e \gamma$, can reach the future MEG sensitivities~\cite{MEG:2016leq,MEGII:2018kmf}.   However, such $\mathcal{O}(1)$ neutrino Yukawa couplings can delay the decoupling of right-handed Dirac neutrinos in the early Universe via, e.g., effective four-fermion interactions mediated by the new scalar, and thus violate the bound of $\Delta N_{eff}$  extracted from the BBN and CMB measurements~\cite{Planck:2018vyg}.  

If the new scalar doublet has instead a keV-scale vacuum, the resulting LFV signals from $\ell_\alpha \to \ell_\beta \gamma$, $\ell_\alpha \to 3\ell_\beta$, $Z\to \ell_\alpha \ell_\beta$, $h\to \ell_\alpha \ell_\beta$, or $\mu-e$ conversion in nuclei would hardly reach the future sensitivities~\cite{Bertuzzo:2015ada}. Nevertheless, such a \textit{flavor-physics inert} model can readily satisfy the thermal conditions predicted by the PTDL mechanism~\cite{Li:2020ner,Li:2021tlv}. In this paper, we will show that, depending on the detailed setup, the model can also manifest itself via a contribution to $\Delta N_{eff}$, which is found to be compatible with the current data, and can be further probed by the forecast sensitivity from, e.g., the CMB Stage-4 (CMB-S4)~\cite{Abazajian:2016yjj} and its combination with the BBN~\cite{Fields:2019pfx}. Thus, $\Delta N_{eff}$ can serve as a promising observable to probe such a keV-vacuum induced Dirac neutrino model, and hence as a complementary avenue to test the PTDL mechanism.
 
The paper is organized as follows. In section~\ref{sec:model}, we recapitulate the  neutrinophilic 2HDM, categorize the two possible thermal conditions in realizing the PTDL mechanism, and then determine the favored vacuum ranges. In section~\ref{sec:evolution}, depending on the thermal conditions, we calculate the respective evolution of the right-handed Dirac neutrinos in the early Universe. The resulting $\Delta N_{eff}$ phenomenology is analyzed in section~\ref{sec:Neff}. Finally, our conclusion is made in section~\ref{sec:conclusion}.

\section{Neutrinophilic 2HDM with a keV-scale vacuum}
\label{sec:model}

The Dirac neutrino masses can be generated by coupling the right-handed Dirac neutrinos to a new Higgs doublet via
\begin{align}\label{nuLag}
-\mathcal{L}_\nu= Y_\nu \bar L \tilde{\Phi} \nu_R+\rm H.c.,
\end{align}
in addition to the SM content. Here, a $Z_2$ (or $U(1)$) symmetry can be imposed to forbid $\nu_R$ from interacting with the SM-like Higgs doublet $\Phi_{\rm SM}$ that is responsible for all the charged fermion masses. In addition, the lepton-number symmetry must be present to forbid the Majorana mass term $\overline{\nu_R^c}\nu_R$, a condition also implemented in the Dirac leptogenesis. The neutrinophilic scalar doublet $\Phi$ is usually assumed to interact with $\Phi_{\rm SM}$ via a $Z_2$- or $U(1)$-symmetric potential, which has a soft-breaking source from $\mu^2 \Phi_{\rm SM}^\dagger \Phi+\rm H.c.$ This soft-breaking term also seeds a seesaw-like suppression of the scalar vacuum $\langle \Phi \rangle =(0,v_\Phi/\sqrt{2})^T$, with~\cite{Gabriel:2006ns,Davidson:2009ha}
\begin{equation}
v_\Phi=\frac{\mu^2 v_{\Phi_{\rm SM}}}{m_{\Phi}^2+\lambda v_{\Phi_{\rm SM}}^2},
\end{equation}
where $(v_{\Phi_{\rm SM}}^2+v_\Phi^2)^{1/2}\simeq 246$~GeV, $m_{\Phi}$ is the bare mass of the neutrinophilic scalar doublet, and $\lambda$ encodes the dimensionless potential parameters. Then, with an electroweak-scale $\Phi$ concerned throughout this paper, a keV-scale vacuum can be readily obtained by $\mu\simeq \mathcal{O}(10^{-2})$~GeV, and hence generates the tiny Dirac neutrino masses via Eq.~\eqref{nuLag}. The resulting hierarchy $v_{\Phi_{\rm SM}}\gg v_\Phi$ suppresses the $\Phi_{\rm SM}$-$\Phi$ mass mixing at the order of $v_\Phi/v_{\Phi_{\rm SM}}\simeq 10^{-8}$, making therefore $\Phi_{\rm SM}$ the SM-like Higgs~\cite{Davidson:2009ha}. It is worth mentioning here that, if the potential exhibits a $U(1)$ symmetry that is only softly broken by the quadratic mixing term, there would be no quartic mixing term, $(\Phi_{\rm SM}^\dagger \Phi)^2+\rm H.c.$, and the neutral components of $\Phi$ would be degenerate in mass~\cite{Machado:2015sha}, which is a distinguishable feature of the $U(1)$-symmetric 2HDM (see, e.g., also Refs.~\cite{Li:2018aov,Li:2020dbg}) and has been considered in the PTDL mechanism~\cite{Li:2021tlv}. Furthermore, given that the constraints from electroweak precision tests already force small mass splitting between the neutral and charged scalars~\cite{Haller:2018nnx}, a strongly first-order phase transition in such a degenerate neutrinophilic 2HDM would be less feasible~\cite{Dorsch:2013wja,Basler:2016obg,Kainulainen:2019kyp} to trigger the electroweak baryogenesis~\cite{Morrissey:2012db}. This leaves the PTDL mechanism~\cite{Li:2020ner,Li:2021tlv} as a natural candidate to address the BAU problem. 
  
\begin{figure*}[t]
  	\centering	
  	\includegraphics[width=0.47\textwidth]{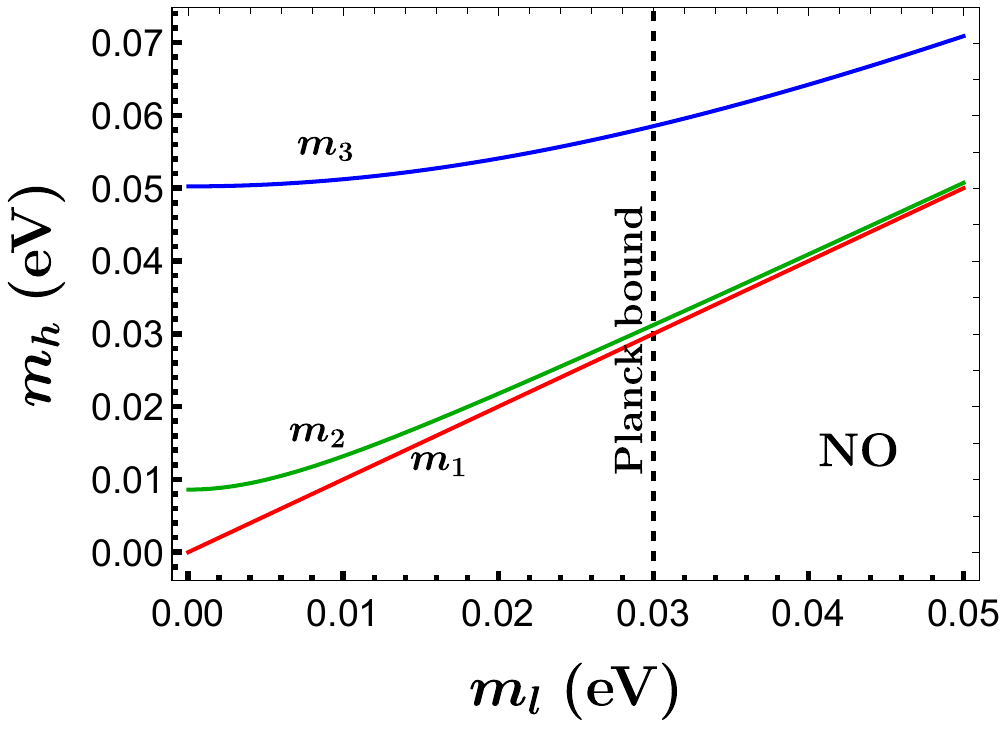}\qquad 	
  	\includegraphics[width=0.47\textwidth]{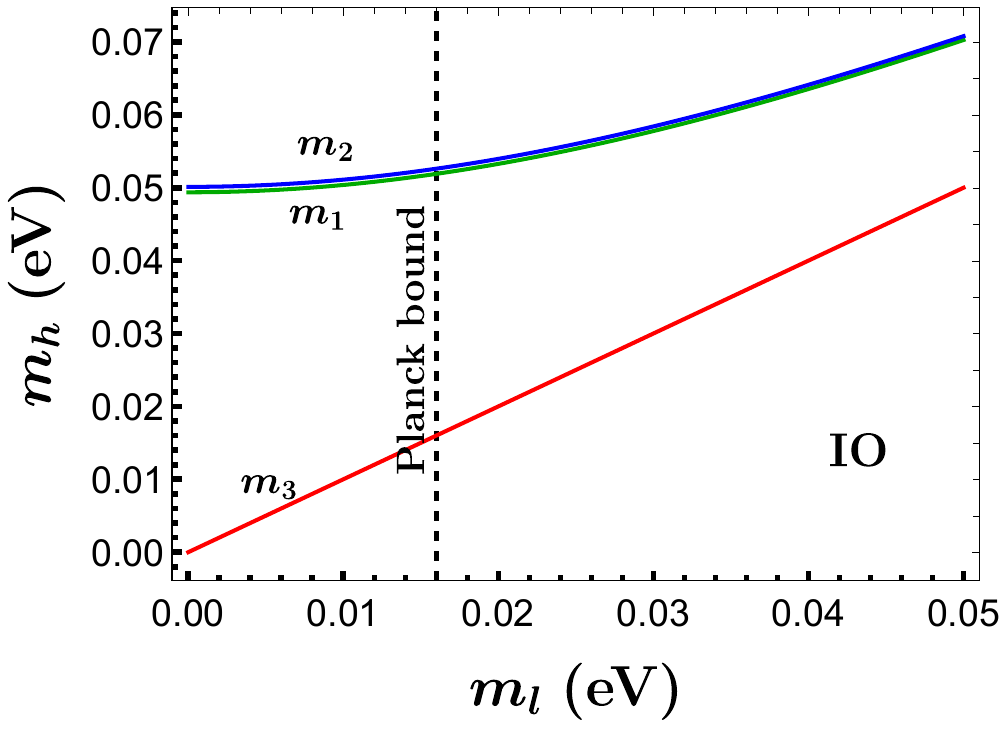}
  	\caption{Current neutrino mass spectrum in the normal-ordering (NO) and inverted-ordering (IO) patterns. The lightest neutrino mass has a maximum value $m_l\simeq 0.030$~eV for the NO and $m_l\simeq 0.016$~eV for the IO pattern after saturating the Planck mass bound $\sum_i m_{\nu_i}<0.12$~eV (vertical line)~\cite{Planck:2018vyg}. We also depict the lightest neutrino mass (red) to have a clear comparison with the two heavier ones (green and blue).} \label{numass} 
\end{figure*}

For the current neutrino masses inferred from oscillation measurements~\cite{Zyla:2020zbs} and cosmological constraints~\cite{Planck:2018vyg}, we expect that at least two neutrinos reside at $0.01-0.05$~eV, as can be clearly seen from Fig.~\ref{numass}, where the neutrino mass spectrum in light of current oscillation data and  under the Planck mass bound $\sum_i m_{\nu_i}<0.12$~eV is depicted. In the neutrinophilic 2HDM framework, such a neutrino mass spectrum indicates a striking property when we consider the evolution of the right-handed Dirac neutrinos in the early Universe. In essence, regardless of the vacuum scales, the two heavier neutrinos ($\nu_{h}$) have a similar order of Yukawa couplings in both the NO and IO patterns. This implies that the two  heavier $\nu_{hR}$ would basically follow the same thermal evolution in the early Universe. On the other hand, given that the lightest neutrino has an upper rather than a lower mass bound, it can either be the case where the lightest neutrino ($\nu_l$) mass is comparable to the two heavier ones, $\mathcal{O}(0.01)$~eV, so that the three $\nu_R$ would exhibit a similar history of evolution, or the case where $\nu_l$ is much lighter so that it would undergo a distinguishable evolution from those of the two heavier ones. Based on the above observations, we can conclude that there are only two possible thermal conditions that would be mostly feasible in realizing the PTDL mechanism in the neutrinophilic 2HDM framework:
\begin{itemize}
 	\item[$(i)$] all the three right-handed Dirac neutrinos never establish thermal equilibrium before the sphaleron decoupling, which is considered in Ref.~\cite{Li:2020ner}.
 	
 	\item[$(ii)$] the lightest $\nu_{lR}$ is out of equilibrium while the other two have already established the left-right equilibration (LRE) prior to the sphaleron freezing out, which is realized and called $\nu_1$-leptogenesis in Ref.~\cite{Li:2021tlv}.
\end{itemize} 
 
Starting firstly with the case $(i)$, we can infer that the late-time LRE requires the decay rate $\Gamma_{\Phi\to \bar L \nu_R} $ to be smaller than the Hubble expansion at the sphaleron decoupling temperature $T_{sph}\approx 132$~GeV~\cite{DOnofrio:2014rug}, which gives
\begin{align}\label{non-therm-i}
	\Gamma_{\Phi\to \bar L \nu_R}  \lesssim 3 H(T)\Big|_{T=T_{sph}},
\end{align} 
where $3H(T)$ comes from the friction term in the Boltzmann equation for the particle-number evolution of $\nu_R$. The Hubble expansion is given by 
\begin{align}
 H(T)\approx 1.66 \sqrt{g_*^\rho}\,T^2 /M_{Pl},
\end{align}
with $g_*^\rho$ the effective number of energy degrees of freedom (d.o.f) and $M_{Pl}\approx 1.2\times 10^{19}$~GeV the Planck mass. Note that the decay rate in Eq.~\eqref{non-therm-i} is given as a sum of three  channels, i.e., one charged and two neutral decay channels for the $\nu_R$ production. Neglecting the charged-lepton and neutrino masses in the final states, the decay rate is given by
\begin{align}
	\Gamma_{\Phi\to \bar L\nu_{iR}}\approx \frac{m_i^2}{4\pi v_\Phi^2}m_\Phi,
\end{align}
where we have replaced the Yukawa coupling $y_{\nu,i}$ by the neutrino mass $m_i$ via $y_{\nu,i}=\sqrt{2}m_i/v_\Phi$, and parametrized the scalar mass at this temperature regime by the free parameter $m_{\Phi}$. Note here that the thermal mass correction of $\Phi$ around the sphaleron decoupling temperature is at $\mathcal{O}(10)$ GeV for small potential parameters $\lambda\simeq \mathcal{O}(0.1)$~\cite{Cline:1995dg}, which ensures that the $m_\Phi$-parametrization is a good approximation when $m_{\Phi}\gtrsim v_{\Phi_{\rm SM}}$. Since Eq.~\eqref{non-therm-i} is applied to three generations of $\nu_R$, requiring the heaviest neutrino ($m_h\simeq 0.05$~eV for both the NO and IO patterns) to satisfy Eq.~\eqref{non-therm-i} would render a lower bound of the vacuum:
\begin{align}\label{vacuum-i}
	\left(\frac{v_\Phi}{\text{keV}}\right)\gtrsim 52\left(\frac{m_{\Phi}}{\text{GeV}}\right)^{1/2},
\end{align}
where $g_*^\rho\approx 106.75$ has been used. Thus, for an electroweak-scale scalar $\Phi$ concerned here, the case $(i)$ necessitates a minimal vacuum around sub-MeV.  

For the case $(ii)$, the production for the two heavier $\nu_{hR}$ should have a rate satisfying 
\begin{align}\label{non-therm-ii}
	\Gamma_{\Phi\to \bar L\nu_{hR}}  \gtrsim 3 H(T)\Big|_{T=T_{sph}}.
\end{align} 
In this case, it is the second lightest neutrino that determines an upper bound of $v_\Phi$. For the IO pattern, since $m_{1,2}\simeq 0.05$~eV, we can immediately obtain as in deriving Eq.~\eqref{vacuum-i} that
\begin{align}
	\left(\frac{v_\Phi}{\text{keV}}\right) \lesssim  52\left(\frac{m_{\Phi}}{\text{GeV}}\right)^{1/2}.  
\end{align}
For the NO pattern, however, $m_2\simeq 0.01$~eV requires the vacuum to be
\begin{align}
	\left(\frac{v_\Phi}{\text{keV}}\right)\lesssim 10 \left(\frac{m_{\Phi}}{\text{GeV}}\right)^{1/2}.  
\end{align} 
Regarding the lightest Dirac neutrino, on the other hand, Eq.~\eqref{non-therm-i} imposes the bound 
\begin{align}\label{non-thermal-ii_1}
	\left(\frac{v_\Phi}{\text{keV}}\right)\gtrsim 1.0 \left(\frac{m_{l}}{\text{meV}}\right)\left(\frac{m_\Phi}{\text{GeV}}\right)^{1/2}.
\end{align}
It should be mentioned that the vacuum scale in case $(ii)$ cannot be arbitrarily low by tuning $m_l$, since  otherwise the two heavier $\nu_{hR}$ will correspond to larger Yukawa couplings, and then the decoupling of $\nu_{hR}$ will be delayed to a sufficiently low temperature via, e.g., the $t$-channel neutrino annihilation, $\nu_{hR} + \bar \nu_{hR} \rightleftharpoons e_L+\bar e_L$. This situation can readily have an impact on the synthesis of primordial elements and the CMB formation. Thus, there actually exists a lower bound of the decoupling temperature for the two heavier $\nu_{hR}$~\cite{Pitrou:2018cgg,Abazajian:2019oqj}, which in turn sets a lower bound on the vacuum. To make this point clear, let us suppose that the late-time decoupling can occur at a lowest allowed temperature $T_{\nu_{hR},dec}\ll m_\Phi$ when the annihilation rate is comparable to  the Hubble expansion. Given that the thermal neutrino annihilation rate $\Gamma_{\nu_{hR},anni}$ scales on naive dimensional grounds as 
\begin{align}
\Gamma_{\nu_{hR},anni}\simeq \frac{m_{h}^4 T^5}{v_\Phi^4m_{\Phi}^4}, 
\end{align}
a lower bound of the vacuum can be obtained as
\begin{align}\label{non-thermal-ii_2}
	\left(\frac{v_\Phi}{\text{keV}}\right)\gtrsim 0.22\left(\frac{m_{h}}{0.01\,\mathrm{eV}}\right) \left[\frac{(T_{\nu_{hR},dec}/\rm GeV)^{3/4}}{m_{\Phi}/\rm GeV}\right].
\end{align}
As $m_{h}\simeq \mathcal{O}(0.01)$~eV for both the NO and IO patterns, it can be seen that the lower bound from Eq.~\eqref{non-thermal-ii_1} would be tighter than from Eq.~\eqref{non-thermal-ii_2} if we focus on the case of $m_l \gtrsim \mathcal{O}(1)$~meV.  

Since the PTDL mechanism considered in Ref.~\cite{Li:2021tlv} favors the NO pattern, the case $(ii)$ would require a vacuum in the following range: 
\begin{align}\label{vacuum-ii}
	1.0\left(\frac{m_{l}}{\text{meV}}\right)\left(\frac{m_\Phi}{\text{GeV}}\right)^{1/2}\lesssim \left(\frac{v_\Phi}{\text{keV}}\right)\lesssim 10\left(\frac{m_{\Phi}}{\text{GeV}}\right)^{1/2},
\end{align}
where $m_l \gtrsim \mathcal{O}(1)$~meV is considered. Together with Eq.~\eqref{vacuum-i}, we can see that a keV-scale or higher vacuum is generically predicted for both cases $(i)$ and $(ii)$. Given that the LFV processes from future sensitivities can only probe a vacuum up to $\mathcal{O}(10)$~eV with an electroweak-scale $\Phi$~\cite{Bertuzzo:2015ada}, there would indeed be rare hope to see the LFV signals in neutrinophilic 2HDM with an $\mathcal{O}(\text{keV})$ vacuum. Nevertheless, the expected sensitivity is dramatically different in cosmic regime, especially given the fact that the current precision of astrophysical and cosmological observations is now making the probe of feeble couplings and light species strikingly possible~\cite{Agrawal:2021dbo}. In the subsequent sections, we will firstly determine the corresponding evolution of $\nu_R$ based on the two cases, and then show that the current limits of $\Delta N_{eff}$ from the BBN, CMB, and their combinations are already available to constrain the $\mathcal{O}(\text{keV})$ vacuum, and the future forecast sensitivity can further test this keV-vacuum induced Dirac neutrino model.

\section{Evolution of right-handed Dirac neutrinos} 
\label{sec:evolution}
   
To estimate the radiation contribution to the SM plasma from $\nu_R$, we now proceed to determine the energy evolution for both cases. For the case $(i)$, since the three $\nu_R$ cannot establish thermalization throughout the Universe expansion, they are essentially produced via the freeze-in mechanism~\cite{Hall:2009bx,Bernal:2017kxu}. Let us consider the energy yield defined by $Y_{\nu_R,\rho}\equiv \rho_{\nu_R}/s_{\rm SM}^{4/3}$, where the SM entropy density is given by 
\begin{align}
s_{\rm SM}=g_*^s \frac{2\pi^2 }{45}T^3, 
\end{align}
with $g_*^s$ denoting the entropy d.o.f. The simplified Boltzmann equation reads
\begin{align}\label{rho-Beq}
	\frac{dY_{\nu_R,\rho}}{dT}=-\frac{C_{\nu_R,\rho}}{s_{\rm SM}^{4/3}H T},
\end{align}
where the collision term $C_{\nu_R,\rho}$  is given by
\begin{align}\label{C-rho_i}
	C_{\nu_R,\rho}
	&=2N_{\nu_R}\int \frac{d^3p_{\Phi}}{(2\pi)^32E_{\Phi}}  f_{\Phi} \int \frac{d^3p_L}{(2\pi)^32E_L} \frac{d^3p_{\nu_R}}{(2\pi)^32E_{\nu_R}} 
	\nonumber \\[0.15cm]
	&\times	(2\pi)^4\delta^4(p_{\Phi}-p_L-p_{\nu_R}) E_{\nu_R}\vert \mathcal{M}_{{\Phi}\to \bar{L} \nu_R} \vert^2.
\end{align}
Here $N_{\nu_R}=6$ if the three $\nu_R$ (and the three $\bar \nu_R$) have similar mass scales so that they have basically the same thermal history, or $N_{\nu_R}=4$ if the lightest $\nu_{lR}$ has a much lower mass scale and hence a negligible effect on the energy budget of the early Universe. The amplitude squared $\vert \mathcal{M}_{{\Phi}\to \bar{L} \nu_R} \vert^2$ sums over the internal d.o.f without average. The factor $2$ in front of the integral in Eq.~\eqref{C-rho_i} simply amounts to the two gauge components of $\Phi$, i.e., here we treat $\Phi$ (as well as $L$) as a single thermal species with two d.o.f. This is because the freeze-in production essentially occurs at the gauge symmetric phase for $m_\Phi\simeq \mathcal{O}(100)$~GeV, and quickly shuts off due to Boltzmann suppression when temperature drops below $m_\Phi$. Note that Eq.~\eqref{C-rho_i} is approximately obtained by neglecting the inverse decay and the Pauli-blocking effects, i.e., by assuming $1-f_{\nu_R,L}\approx 1$. With the Boltzmann distribution $f_\Phi=e^{-E/T}$, the energy density at decoupling is then given by
\begin{align}\label{rho_i}
\rho_{\nu_R, dec}
	&\approx s_{\rm SM}^{4/3}(T_{\nu_R, dec})\int_{0}^{\infty}\frac{C_{\nu_R,\rho}}{s_{\rm SM}^{4/3}H T}dT
	\nonumber \\[0.15cm]
	&\approx   0.09 N_{\nu_R} \left(\frac{106.75}{g_{\nu_R, dec}}\right)^{1/2} T_{\nu_R, dec}^4
	\nonumber \\[0.15cm]
	&\times \left(\frac{m_{\nu} }{0.01\,\rm eV}\right)^2 
 \left(\frac{100\,\text{GeV}}{m_\Phi}\right)\left(\frac{500\,\text{keV}}{v_\Phi}\right)^2,
\end{align}
where we have taken the approximation $g_*^\rho\approx g_*^s\approx g_{\nu_R,dec}$ determined at the neutrino freeze-in temperature $T_{\nu_R, dec}$. After the decoupling of $\nu_R$, the energy exchange between the SM plasma and the right-handed Dirac neutrinos ceases, and
the energy density $\rho_{\nu_R, ref}$ at a late-time reference temperature $T_{ref}$ can be readily rescaled by 
\begin{align}\label{late-time rho}
	\frac{\rho_{\nu_{R}, ref}}{\rho_{\nu_{R}, dec}}=\frac{\rho_{\text{SM}, ref}}{\rho_{\text{SM}, dec}},
\end{align}
where the SM energy density is given by 
\begin{align}
\rho_{\text{SM}}=g_*^\rho \frac{\pi^2}{30} T^4.
\end{align}

For the case $(ii)$, on the other hand, the energy density from the thermalized $\nu_{h R}$ after freezing out is given by Eq.~\eqref{late-time rho}, with a thermal energy spectrum
\begin{align}\label{nuhR rho}
	\rho_{\nu_{hR}}= N_{\nu_{hR}}\frac{7\pi^2}{240}~T_{\nu_{hR}}^4,
\end{align}
where $N_{\nu_{h R}}=4$ takes into account the two heavier $\nu_{hR}$ and their antiparticles, and $T_{\nu_{hR}}$ denotes the $\nu_{hR}$ temperature after freezing out. For the lightest neutrino $\nu_{lR}$ in case $(ii)$, the energy evolution follows Eqs.~\eqref{rho-Beq}--\eqref{rho_i}, with $N_{\nu_R}=2$.

\section{Promising signals from effective neutrino number shift} 
\label{sec:Neff}

The extra radiation contributing to the SM plasma in the early Universe can be parametrized by the shift of the effective neutrino number via 
\begin{align}
	\Delta N_{eff}\equiv  \frac{\rho_{rad}}{\rho_{\nu_L}},
\end{align}
where $\rho_{\nu_L}$ is the energy density of a left-handed neutrino. In the SM, $N_{eff}^{\rm SM}=3$ just prior to the BBN and $N_{eff}^{\rm SM} = 3.044-3.045$~\cite{Mangano:2005cc,deSalas:2016ztq,Gariazzo:2019gyi,Escudero:2020dfa,Akita:2020szl,Froustey:2020mcq,Bennett:2020zkv} after taking into account the non-instantaneous decoupling of active neutrinos below $T=\mathcal{O}(1)$~MeV. The extra radiation contribution to $\Delta N_{eff}$ can also be expressed in terms of the SM energy density via~\cite{Luo:2020fdt,Li:2021okx}
\begin{align}\label{Neff def}
	\Delta N_{eff} = \frac{4}{7} \left[\frac{g_{*}^s(T_{\gamma e \nu})}{g_{*}^s(T_{ref})}\right]^{4/3} g_*^\rho (T_{ref})\, \frac{\rho_{rad}(T_{ref})}{\rho_{\text{SM}}(T_{ref})},
\end{align}
where $g_{*}^s(T_{\gamma e \nu})=10.75$ corresponds to the epoch when the relativistic SM plasma contains photons, electrons, positrons, and neutrinos. 

According to Eq.~\eqref{Neff def}, we can calculate the shift $\Delta N_{eff}$ at the decoupling temperature $T_{ref}=T_{\nu_R, dec}$ in case $(i)$, giving
\begin{align}\label{Neff def-i}
\Delta N_{eff}^{(i)}&\approx 7.345\times 10^{-3}N_{\nu_R} 
\nonumber \\[0.15cm]
&\times \left(\frac{m_{\nu} }{0.01\,\rm eV}\right)^2
 \left(\frac{100\,\text{GeV}}{m_\Phi}\right)\left(\frac{500\,\text{keV}}{v_\Phi}\right)^2,
\end{align}
where $g_{\nu_R, dec}\approx 106.75$ has been used since we expect that $T_{\nu_R, dec}\simeq \mathcal{O}(m_{\Phi})$. For the case $(ii)$, we can determine the shift $\Delta N_{eff}$ from the two heavier $\nu_{hR}$ by applying Eq.~\eqref{nuhR rho} to Eq.~\eqref{Neff def} at $T_{ref}=T_{\nu_{hR}, dec}$, while the shift caused by $\nu_{lR}$ is given by Eq.~\eqref{Neff def-i} with $N_{\nu_R}=2$. The resulting total shift in case $(ii)$ is then given by
\begin{align}
\Delta N_{eff}^{(ii)}&\approx 0.0037 \left(\frac{m_{l} }{\rm meV}\right)^2
\left(\frac{100\,\text{GeV}}{m_\Phi}\right)\left(\frac{100\,\text{keV}}{v_\Phi}\right)^2
\nonumber \\[0.15cm]
& + 0.0937 \left(\frac{106.75}{g_{\nu_{hR},dec}}\right)^{4/3}.
\end{align}
For an electroweak-scale decoupling temperature of the heavier neutrinos, we have $g_{\nu_{h R},dec}=106.75$, the maximum amount of entropy d.o.f available from SM particles.

Currently, a combined constraint from BBN (including primordial abundances of helium-4 and deuterium) and CMB, i.e., CMB+BBN+$Y_p$+D, gives $N_{eff}=2.843\pm 0.154$, setting an upper limit $\Delta N_{eff}=N_{eff}-3<0.151$ at the $2\sigma$ level~\cite{Fields:2019pfx}, while the severest bound from the Planck 2018 results is given by $N_{eff}=2.99\pm 0.17$, limiting $\Delta N_{eff}=N_{eff}-3.045<0.285$ at the $95\%$ confidence level~\cite{Planck:2018vyg}.   The future forecast sensitivity from CMB-S4 can reach $\Delta N_{eff}=\mathcal{O}(0.01)$, depending on the sky fraction $f_{sky}$~\cite{Abazajian:2016yjj}. In addition, the forecast of $N_{eff}$ from BBN+CMB-S4 can reach a $1\sigma$ sensitivity, $\sigma_{S4}(N_{eff}|\text{BBN})=0.030$~\cite{Fields:2019pfx}. On the other hand, the South Pole Telescope SPT-3G is expected to have a $2\sigma$ sensitivity of $\Delta N_{eff}<0.116$~\cite{Benson:2014qhw}. Given that the bound from CMB+BBN+$Y_p$+D is tighter than the Planck measurements, we will use the former as a constraint and apply the forecast sensitivities from SPT-3G and $\sigma_{S4}(N_{eff}|\text{BBN})$ to probe the vacuum scale. 
Noticeably, since $\Delta N_{eff}^{(ii)}\gtrsim 0.0937$, which corresponds to an electroweak-scale $T_{\nu_{hR}, dec}$ and a negligible effect from the lightest $\nu_{lR}$, the future BBN+CMB-S4 is able to exclude this second possibility of the PTDL mechanism, or the $\nu_1$-leptogenesis~\cite{Li:2021tlv}.

\begin{figure}[t]
	\centering	
	\includegraphics[width=0.47\textwidth]{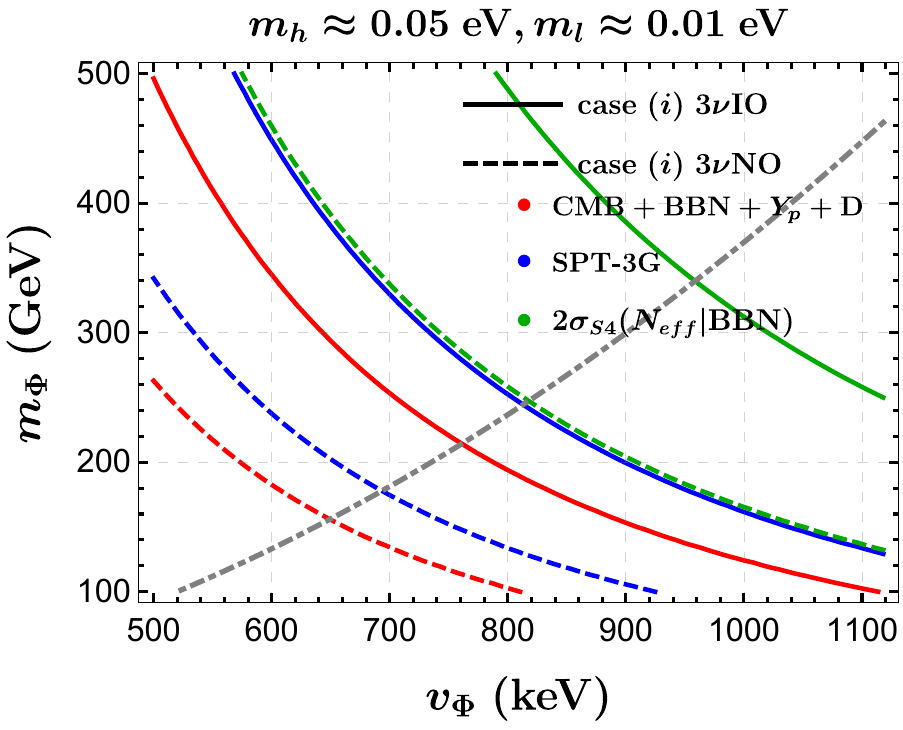} 
	\caption{Correlation between the electroweak-scale scalar mass $m_{\Phi}$ and the vacuum  scale $v_{\Phi}$ in case $(i)$, by confronting the current CMB+BBN+$Y_p$+D bound (red) as well as the forecast sensitivities from SPT-3G (blue) and BBN+CMB-S4 (green), all being at the $2\sigma$ level, to the resulting shift $\Delta N_{eff}^{(i)}$ from the three $\nu_R$ for both the NO (dashed curves) and IO (solid curves) patterns, where the heavier neutrino mass is set at $m_h\approx 0.05$~eV and the lighter one at $m_l\approx 0.01$~eV, respectively. The regions below the different curves are already excluded by the corresponding constraints, while the region below the gray dash-dotted curve represents the thermal condition required by Eq.~\eqref{vacuum-i}.} \label{Vac1} 
\end{figure}

To estimate the maximal $\Delta N_{eff}^{(i)}$ that allows an observable imprint from the lightest neutrino, we assume that its mass $m_{l}$ reaches already the order of the Planck bound, i.e., $m_{l}\simeq \mathcal{O}(0.01)$~eV for both the NO and IO patterns (see also Fig.~\ref{numass}), so that all the three $\nu_R$ have comparable contributions to $\Delta N_{eff}$. Specifically, for the NO pattern, $\Delta N_{eff}^{(i)}$ is induced by one heavy $\nu_R$ with $m_{3}\approx 0.05$~eV ($N_{\nu_R}=2$) and two lighter $\nu_R$ with  $m_{1,2}\approx 0.01$~eV ($N_{\nu_R}=4$), while for the IO pattern, $\Delta N_{eff}^{(i)}$  comes from two heavier $\nu_R$ with $m_{1,2}\approx 0.05$~eV ($N_{\nu_R}=4$) and the lightest $\nu_R$ with $m_{3}\approx 0.01$~eV ($N_{\nu_R}=2$). Confronting the current CMB+BBN+$Y_p$+D bound (red) as well as the forecast sensitivities from SPT-3G (blue) and BBN+CMB-S4 (green), all being at the $2\sigma$ level, to the resulting shift $\Delta N_{eff}^{(i)}$ from the three $\nu_R$ for both the NO (dashed curves) and IO (solid curves) patterns, we show in Fig.~\ref{Vac1} the correlation between the electroweak-scale scalar mass $m_{\Phi}$ and the vacuum scale $v_{\Phi}$ in case $(i)$, where the heavier neutrino mass is set at $m_h\approx 0.05$~eV and the lighter one at $m_l\approx 0.01$~eV, respectively. 

As can be seen from Fig.~\ref{Vac1}, the vacuum in case $(i)$ is generically pushed up to $\mathcal{O}(1)$~MeV for $m_\Phi\simeq \mathcal{O}(100)$~GeV. For example, a possible excess of the effective neutrino number around $2\sigma_{S4}(N_{eff}|\text{BBN})=0.06$ can be generated by a vacuum at $1$~MeV with $m_\Phi\approx 300$~GeV, which can be tested by the future BBN+CMB-S4 sensitivity. For a much higher vacuum scale, however, the resulting shift $\Delta N_{eff}^{(i)}$ becomes negligible, as can be seen from Eq.~\eqref{Neff def-i}. It should be mentioned that, just as the case $(i)$ realized in Ref.~\cite{Li:2020ner} via a Yukawa texture-dependent BAU, we have shown there that the required vacuum is predicted to be $\mathcal{O}(0.1)$~GeV in the neutrinophilic 2HDM. In this case, it would be hardly possible to test the PTDL mechanism if the Yukawa structures presumed in Ref.~\cite{Li:2020ner} are indeed responsible for the observed neutrino mixing and masses.

For an estimate of the shift $\Delta N_{eff}^{(ii)}$, it should be born in mind that the lightest neutrino in the NO pattern cannot reach $\mathcal{O}(0.01)$~eV, because otherwise the out-of-equilibrium condition would be violated, as can also be seen from Eq.~\eqref{vacuum-ii}. If the lightest neutrino has a much smaller mass, on the other hand, the shift $\Delta N_{eff}^{(ii)}$ would primarily come from the two thermalized $\nu_{hR}$, and hence depend on the decoupling temperature $T_{\nu_{hR},dec}$. This has been recently analyzed in Ref.~\cite{Abazajian:2019oqj}, pointing out that $T_{\nu_{hR},dec}$ must be higher than the QCD phase transition temperature under the current limit from Planck 2018 release and will be pushed to $\mathcal{O}(10)$~GeV by the future SPT-3G sensitivity. For our consideration here, we take $m_{l}=1$~meV as a benchmark scale to include a non-negligible contribution from the freeze-in $\nu_{lR}$. Note that, given the thermal condition presented in Eq.~\eqref{vacuum-ii}, we can determine the range of $m_\Phi$ in terms of $m_l$ and $v_\Phi$, which, after fixing $m_l=1$~meV, would translate into an interval of $\Delta N_{eff}^{(ii)}$ in terms of the vacuum $v_\Phi$ and the d.o.f $g_{\nu_{hR},dec}$. 

\begin{figure}[t]
	\centering	
	\includegraphics[width=0.48\textwidth]{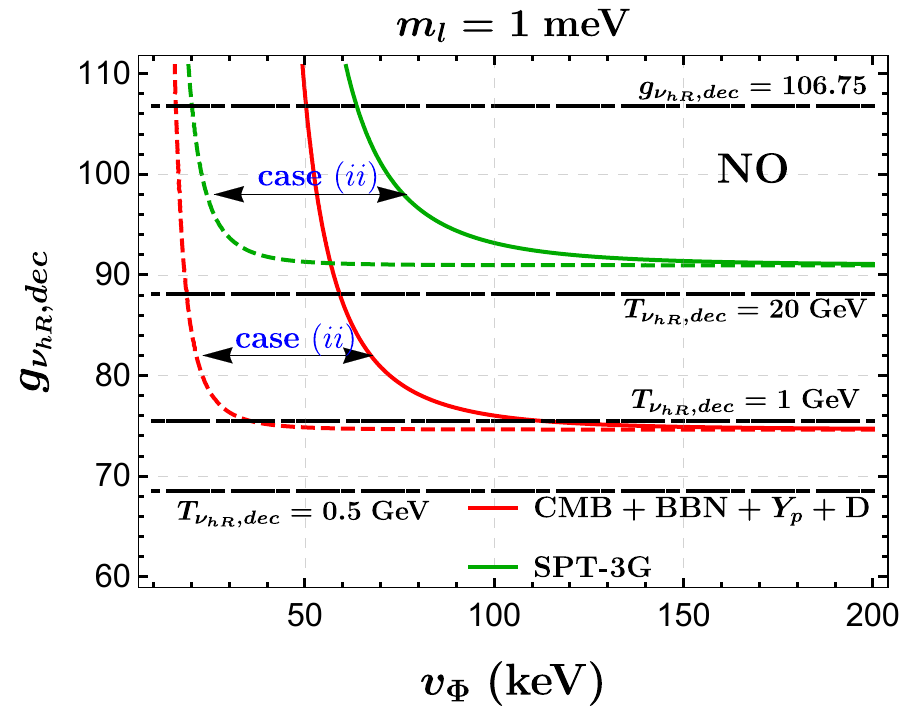}
	\caption{The vacuum scale $v_{\Phi}$ in case $(ii)$ with a varying $T_{\nu_{hR},dec}$ in the NO pattern, as constrained by the current CMB+BBN+$Y_p$+D (red) and will be probed by the future SPT-3G (green) sensitivities. The otherwise unknown scalar mass $m_\Phi$ has been translated into an interval between the solid (upper bound) and dashed (lower bound) curves as governed by the thermal condition of Eq.~\eqref{vacuum-ii}, by fixing $m_l=1$~meV.} \label{Vac2} 
\end{figure}
In Fig.~\ref{Vac2}, the solid (dashed) curve corresponds to the upper (lower) bound of  $m_\Phi$ governed by the thermal condition of Eq.~\eqref{vacuum-ii}. Under this prescription, it can be seen that the vacuum in the range $30\lesssim v_\Phi/\text{keV}\lesssim 110$ with a decoupling temperature at $1$~GeV can generate a shift $\Delta N_{eff}$ of the current CMB+BBN+$Y_p$+D level. It is also found that, under the CMB+BBN+$Y_p$+D bound, the decoupling temperature $T_{\nu_{hR},dec}$ must be larger than $0.5$~GeV, independent of the keV-scale vacuum considered. In light of the future SPT-3G sensitivity, if no excess at the level of $\Delta N_{eff}=0.116$ is observed, the lower bound of the decoupling temperature would be pushed up to $20$~GeV. For $g_{\nu_{hR},dec}=106.75$ corresponding to an electroweak freezing-out temperature of $\nu_{hR}$, a vacuum in the range $20\lesssim v_\Phi/\text{keV}\lesssim 65$ can generate a shift $\Delta N_{eff}$ of the future SPT-3G level. As mentioned before, the future BBN+CMB-S4 can readily exclude the case $(ii)$, since a minimal value $\Delta N_{eff}^{(ii)}\simeq 0.0937$ is expected in this case.

Finally, it is interesting to point out that, for a vacuum being around $100$~keV, the lightest Dirac neutrino with a mass of $\mathcal{O}(1)$~meV will lead to a Yukawa coupling of $\mathcal{O}(10^{-8})$, and the two heavier ones have the electronic Yukawa of $\mathcal{O}(10^{-6})$. In this respect, the neutrinophilic 2HDM that explains the Dirac neutrino masses and the BAU problem does not deteriorate too much the Yukawa hierarchy and naturalness issue encountered already in the SM, which is different from the eV-scale vacuum where the neutrino Yukawa couplings are naturally of $\mathcal{O}(1)$, while the light charged fermion Yukawas deviate significantly from $\mathcal{O}(1)$. We may then expect a unified framework to explain the Yukawa feebleness in such a keV-vacuum Dirac neutrino model.

\section{Conclusion}
\label{sec:conclusion}

The fact that no observation as yet of Dirac neutrino related signals in both direct collider detection and indirect low-energy flavor physics could just hint towards a feeble Dirac neutrino Yukawa regime, a similar pattern that already existed in the charged fermion Yukawa sector of the SM. We have shown in this paper that, while the neutrinophilic 2HDM with a keV-scale vacuum is inert in low-energy flavor physics such that the observation of LFV processes cannot be expected in future experiments, the sensitivities from cosmic regime are able to probe such a \textit{flavor-physics inert} model. Besides being distinguishable from the lighter-vacuum cases with observable LFV processes, the relativistic right-handed Dirac neutrinos contribute to the energy budget of the early SM plasma, prompting significant shift of the effective neutrino number. The current measurement from CMB+BBN+$Y_p$+D has already presented a restrictive regime, and the future forecast from SPT-3G and BBN+CMB-S4 is able to test or even fully exclude the keV-vacuum Dirac neutrino model, in which the BAU enigma could be successfully solved by the PTDL mechanism.   

\section*{Acknowledgments}
This work is supported by the National Natural Science Foundation of China under Grant Nos.~12135006, 12075097, 12047527, and 11775092, as well as by the Fundamental Research Funds for the Central Universities under Grant No.~CCNU20TS007.

\bibliographystyle{JHEP}
\bibliography{reference}

\providecommand{\href}[2]{#2}\begingroup\raggedright\begin{thebibliography}{10}

\bibitem{Dick:1999je}
K.~Dick, M.~Lindner, M.~Ratz, and D.~Wright, {\it {Leptogenesis with Dirac
  neutrinos}},  {\it Phys. Rev. Lett.} {\bf 84} (2000) 4039--4042,
  [\href{http://arxiv.org/abs/hep-ph/9907562}{{\tt hep-ph/9907562}}].

\bibitem{Murayama:2002je}
H.~Murayama and A.~Pierce, {\it {Realistic Dirac leptogenesis}},  {\it Phys.
  Rev. Lett.} {\bf 89} (2002) 271601,
  [\href{http://arxiv.org/abs/hep-ph/0206177}{{\tt hep-ph/0206177}}].

\bibitem{Li:2020ner}
S.-P. Li, X.-Q. Li, X.-S. Yan, and Y.-D. Yang, {\it {Freeze-in Dirac
  neutrinogenesis: thermal leptonic CP asymmetry}},  {\it Eur. Phys. J. C} {\bf
  80} (2020), no.~12 1122, [\href{http://arxiv.org/abs/2005.02927}{{\tt
  arXiv:2005.02927}}].

\bibitem{Li:2021tlv}
S.-P. Li, X.-Q. Li, X.-S. Yan, and Y.-D. Yang, {\it {Baryogenesis from
  hierarchical Dirac neutrinos}},  {\it Phys. Rev. D} {\bf 104} (2021), no.~11
  115014, [\href{http://arxiv.org/abs/2105.01317}{{\tt arXiv:2105.01317}}].

\bibitem{Steigman:2012ve}
G.~Steigman, {\it {Neutrinos And Big Bang Nucleosynthesis}},  {\it Adv. High
  Energy Phys.} {\bf 2012} (2012) 268321,
  [\href{http://arxiv.org/abs/1208.0032}{{\tt arXiv:1208.0032}}].

\bibitem{Cyburt:2015mya}
R.~H. Cyburt, B.~D. Fields, K.~A. Olive, and T.-H. Yeh, {\it {Big Bang
  Nucleosynthesis: 2015}},  {\it Rev. Mod. Phys.} {\bf 88} (2016) 015004,
  [\href{http://arxiv.org/abs/1505.01076}{{\tt arXiv:1505.01076}}].

\bibitem{Pitrou:2018cgg}
C.~Pitrou, A.~Coc, J.-P. Uzan, and E.~Vangioni, {\it {Precision big bang
  nucleosynthesis with improved Helium-4 predictions}},  {\it Phys. Rept.} {\bf
  754} (2018) 1--66, [\href{http://arxiv.org/abs/1801.08023}{{\tt
  arXiv:1801.08023}}].

\bibitem{Fields:2019pfx}
B.~D. Fields, K.~A. Olive, T.-H. Yeh, and C.~Young, {\it {Big-Bang
  Nucleosynthesis after Planck}},  {\it JCAP} {\bf 03} (2020) 010,
  [\href{http://arxiv.org/abs/1912.01132}{{\tt arXiv:1912.01132}}]. [Erratum:
  JCAP 11, E02 (2020)].

\bibitem{Planck:2018vyg}
{\bf Planck} Collaboration, N.~Aghanim et~al., {\it {Planck 2018 results. VI.
  Cosmological parameters}},  {\it Astron. Astrophys.} {\bf 641} (2020) A6,
  [\href{http://arxiv.org/abs/1807.06209}{{\tt arXiv:1807.06209}}]. [Erratum:
  Astron.Astrophys. 652, C4 (2021)].

\bibitem{Abazajian:2019oqj}
K.~N. Abazajian and J.~Heeck, {\it {Observing Dirac neutrinos in the cosmic
  microwave background}},  {\it Phys. Rev. D} {\bf 100} (2019) 075027,
  [\href{http://arxiv.org/abs/1908.03286}{{\tt arXiv:1908.03286}}].

\bibitem{Luo:2020sho}
X.~Luo, W.~Rodejohann, and X.-J. Xu, {\it {Dirac neutrinos and $N_{{\rm
  eff}}$}},  {\it JCAP} {\bf 06} (2020) 058,
  [\href{http://arxiv.org/abs/2005.01629}{{\tt arXiv:2005.01629}}].

\bibitem{Adshead:2020ekg}
P.~Adshead, Y.~Cui, A.~J. Long, and M.~Shamma, {\it {Unraveling the Dirac
  neutrino with cosmological and terrestrial detectors}},  {\it Phys. Lett. B}
  {\bf 823} (2021) 136736, [\href{http://arxiv.org/abs/2009.07852}{{\tt
  arXiv:2009.07852}}].

\bibitem{Luo:2020fdt}
X.~Luo, W.~Rodejohann, and X.-J. Xu, {\it {Dirac neutrinos and N$_{eff}$.
  Part~II. The freeze-in case}},  {\it JCAP} {\bf 03} (2021) 082,
  [\href{http://arxiv.org/abs/2011.13059}{{\tt arXiv:2011.13059}}].

\bibitem{Gabriel:2006ns}
S.~Gabriel and S.~Nandi, {\it {A New two Higgs doublet model}},  {\it Phys.
  Lett. B} {\bf 655} (2007) 141--147,
  [\href{http://arxiv.org/abs/hep-ph/0610253}{{\tt hep-ph/0610253}}].

\bibitem{Davidson:2009ha}
S.~M. Davidson and H.~E. Logan, {\it {Dirac neutrinos from a second Higgs
  doublet}},  {\it Phys. Rev. D} {\bf 80} (2009) 095008,
  [\href{http://arxiv.org/abs/0906.3335}{{\tt arXiv:0906.3335}}].

\bibitem{Atwood:2005bf}
D.~Atwood, S.~Bar-Shalom, and A.~Soni, {\it {Neutrino masses, mixing and
  leptogenesis in a two Higgs doublet model 'for the third generation'}},  {\it
  Phys. Lett. B} {\bf 635} (2006) 112--117,
  [\href{http://arxiv.org/abs/hep-ph/0502234}{{\tt hep-ph/0502234}}].

\bibitem{Clarke:2015hta}
J.~D. Clarke, R.~Foot, and R.~R. Volkas, {\it {Natural leptogenesis and
  neutrino masses with two Higgs doublets}},  {\it Phys. Rev. D} {\bf 92}
  (2015), no.~3 033006, [\href{http://arxiv.org/abs/1505.05744}{{\tt
  arXiv:1505.05744}}].

\bibitem{Li:2018rax}
S.-P. Li, X.-Q. Li, Y.-D. Yang, and X.~Zhang, {\it {$
  {R}_{D^{\left(*\right)}},{R}_{K^{\left(*\right)}} $ and neutrino mass in the
  2HDM-III with right-handed neutrinos}},  {\it JHEP} {\bf 09} (2018) 149,
  [\href{http://arxiv.org/abs/1807.08530}{{\tt arXiv:1807.08530}}].

\bibitem{Li:2019xmi}
S.-P. Li and X.-Q. Li, {\it {Probing new physics signals with symmetry-restored
  Yukawa textures}},  {\it Eur. Phys. J. C} {\bf 80} (2020), no.~3 268,
  [\href{http://arxiv.org/abs/1907.13555}{{\tt arXiv:1907.13555}}].

\bibitem{Cogollo:2019mbd}
D.~Cogollo, R.~D. Matheus, T.~B. de~Melo, and F.~S. Queiroz, {\it {Type I + II
  Seesaw in a Two Higgs Doublet Model}},  {\it Phys. Lett. B} {\bf 797} (2019)
  134813, [\href{http://arxiv.org/abs/1904.07883}{{\tt arXiv:1904.07883}}].

\bibitem{Davidson:2010sf}
S.~M. Davidson and H.~E. Logan, {\it {LHC phenomenology of a two-Higgs-doublet
  neutrino mass model}},  {\it Phys. Rev. D} {\bf 82} (2010) 115031,
  [\href{http://arxiv.org/abs/1009.4413}{{\tt arXiv:1009.4413}}].

\bibitem{Machado:2015sha}
P.~A.~N. Machado, Y.~F. Perez, O.~Sumensari, Z.~Tabrizi, and R.~Z. Funchal,
  {\it {On the Viability of Minimal Neutrinophilic Two-Higgs-Doublet Models}},
  {\it JHEP} {\bf 12} (2015) 160, [\href{http://arxiv.org/abs/1507.07550}{{\tt
  arXiv:1507.07550}}].

\bibitem{Bertuzzo:2015ada}
E.~Bertuzzo, Y.~F. Perez~G., O.~Sumensari, and R.~Zukanovich~Funchal, {\it
  {Limits on Neutrinophilic Two-Higgs-Doublet Models from Flavor Physics}},
  {\it JHEP} {\bf 01} (2016) 018, [\href{http://arxiv.org/abs/1510.04284}{{\tt
  arXiv:1510.04284}}].

\bibitem{MEG:2016leq}
{\bf MEG} Collaboration, A.~M. Baldini et~al., {\it {Search for the lepton
  flavour violating decay $\mu ^+ \rightarrow \mathrm {e}^+ \gamma $ with the
  full dataset of the MEG experiment}},  {\it Eur. Phys. J. C} {\bf 76} (2016),
  no.~8 434, [\href{http://arxiv.org/abs/1605.05081}{{\tt arXiv:1605.05081}}].

\bibitem{MEGII:2018kmf}
{\bf MEG II} Collaboration, A.~M. Baldini et~al., {\it {The design of the MEG
  II experiment}},  {\it Eur. Phys. J. C} {\bf 78} (2018), no.~5 380,
  [\href{http://arxiv.org/abs/1801.04688}{{\tt arXiv:1801.04688}}].

\bibitem{Abazajian:2016yjj}
{\bf CMB-S4} Collaboration, K.~N. Abazajian et~al., {\it {CMB-S4 Science Book,
  First Edition}},  \href{http://arxiv.org/abs/1610.02743}{{\tt
  arXiv:1610.02743}}.

\bibitem{Li:2018aov}
S.-P. Li, X.-Q. Li, and Y.-D. Yang, {\it {Muon $g-2$ in a $U(1)$-symmetric
  Two-Higgs-Doublet Model}},  {\it Phys. Rev. D} {\bf 99} (2019), no.~3 035010,
  [\href{http://arxiv.org/abs/1808.02424}{{\tt arXiv:1808.02424}}].

\bibitem{Li:2020dbg}
S.-P. Li, X.-Q. Li, Y.-Y. Li, Y.-D. Yang, and X.~Zhang, {\it {Power-aligned
  2HDM: a correlative perspective on $(g-2)_{e,\mu}$}},  {\it JHEP} {\bf 01}
  (2021) 034, [\href{http://arxiv.org/abs/2010.02799}{{\tt arXiv:2010.02799}}].

\bibitem{Haller:2018nnx}
J.~Haller, A.~Hoecker, R.~Kogler, K.~M\"onig, T.~Peiffer, and J.~Stelzer, {\it
  {Update of the global electroweak fit and constraints on two-Higgs-doublet
  models}},  {\it Eur. Phys. J. C} {\bf 78} (2018), no.~8 675,
  [\href{http://arxiv.org/abs/1803.01853}{{\tt arXiv:1803.01853}}].

\bibitem{Dorsch:2013wja}
G.~Dorsch, S.~Huber, and J.~No, {\it {A strong electroweak phase transition in
  the 2HDM after LHC8}},  {\it JHEP} {\bf 10} (2013) 029,
  [\href{http://arxiv.org/abs/1305.6610}{{\tt arXiv:1305.6610}}].

\bibitem{Basler:2016obg}
P.~Basler, M.~Krause, M.~Muhlleitner, J.~Wittbrodt, and A.~Wlotzka, {\it
  {Strong First Order Electroweak Phase Transition in the CP-Conserving 2HDM
  Revisited}},  {\it JHEP} {\bf 02} (2017) 121,
  [\href{http://arxiv.org/abs/1612.04086}{{\tt arXiv:1612.04086}}].

\bibitem{Kainulainen:2019kyp}
K.~Kainulainen, V.~Keus, L.~Niemi, K.~Rummukainen, T.~V.~I. Tenkanen, and
  V.~Vaskonen, {\it {On the validity of perturbative studies of the electroweak
  phase transition in the Two Higgs Doublet model}},  {\it JHEP} {\bf 06}
  (2019) 075, [\href{http://arxiv.org/abs/1904.01329}{{\tt arXiv:1904.01329}}].

\bibitem{Morrissey:2012db}
D.~E. Morrissey and M.~J. Ramsey-Musolf, {\it {Electroweak baryogenesis}},
  {\it New J. Phys.} {\bf 14} (2012) 125003,
  [\href{http://arxiv.org/abs/1206.2942}{{\tt arXiv:1206.2942}}].

\bibitem{Zyla:2020zbs}
{\bf Particle Data Group} Collaboration, P.~A. Zyla et~al., {\it {Review of
  Particle Physics}},  {\it PTEP} {\bf 2020} (2020), no.~8 083C01 and 2021
  update.

\bibitem{DOnofrio:2014rug}
M.~D'Onofrio, K.~Rummukainen, and A.~Tranberg, {\it {Sphaleron Rate in the
  Minimal Standard Model}},  {\it Phys. Rev. Lett.} {\bf 113} (2014), no.~14
  141602, [\href{http://arxiv.org/abs/1404.3565}{{\tt arXiv:1404.3565}}].

\bibitem{Cline:1995dg}
J.~M. Cline, K.~Kainulainen, and A.~P. Vischer, {\it {Dynamics of two Higgs
  doublet CP violation and baryogenesis at the electroweak phase transition}},
  {\it Phys. Rev. D} {\bf 54} (1996) 2451--2472,
  [\href{http://arxiv.org/abs/hep-ph/9506284}{{\tt hep-ph/9506284}}].

\bibitem{Agrawal:2021dbo}
P.~Agrawal et~al., {\it {Feebly-interacting particles: FIPs 2020 workshop
  report}},  {\it Eur. Phys. J. C} {\bf 81} (2021), no.~11 1015,
  [\href{http://arxiv.org/abs/2102.12143}{{\tt arXiv:2102.12143}}].

\bibitem{Hall:2009bx}
L.~J. Hall, K.~Jedamzik, J.~March-Russell, and S.~M. West, {\it {Freeze-In
  Production of FIMP Dark Matter}},  {\it JHEP} {\bf 03} (2010) 080,
  [\href{http://arxiv.org/abs/0911.1120}{{\tt arXiv:0911.1120}}].

\bibitem{Bernal:2017kxu}
N.~Bernal, M.~Heikinheimo, T.~Tenkanen, K.~Tuominen, and V.~Vaskonen, {\it {The
  Dawn of FIMP Dark Matter: A Review of Models and Constraints}},  {\it Int. J.
  Mod. Phys. A} {\bf 32} (2017), no.~27 1730023,
  [\href{http://arxiv.org/abs/1706.07442}{{\tt arXiv:1706.07442}}].

\bibitem{Mangano:2005cc}
G.~Mangano, G.~Miele, S.~Pastor, T.~Pinto, O.~Pisanti, and P.~D. Serpico, {\it
  {Relic neutrino decoupling including flavor oscillations}},  {\it Nucl. Phys.
  B} {\bf 729} (2005) 221--234,
  [\href{http://arxiv.org/abs/hep-ph/0506164}{{\tt hep-ph/0506164}}].

\bibitem{deSalas:2016ztq}
P.~F. de~Salas and S.~Pastor, {\it {Relic neutrino decoupling with flavour
  oscillations revisited}},  {\it JCAP} {\bf 07} (2016) 051,
  [\href{http://arxiv.org/abs/1606.06986}{{\tt arXiv:1606.06986}}].

\bibitem{Gariazzo:2019gyi}
S.~Gariazzo, P.~F. de~Salas, and S.~Pastor, {\it {Thermalisation of sterile
  neutrinos in the early Universe in the 3+1 scheme with full mixing matrix}},
  {\it JCAP} {\bf 07} (2019) 014, [\href{http://arxiv.org/abs/1905.11290}{{\tt
  arXiv:1905.11290}}].

\bibitem{Escudero:2020dfa}
M.~Escudero~Abenza, {\it {Precision early universe thermodynamics made simple:
  $N_{\rm eff}$ and neutrino decoupling in the Standard Model and beyond}},
  {\it JCAP} {\bf 05} (2020) 048, [\href{http://arxiv.org/abs/2001.04466}{{\tt
  arXiv:2001.04466}}].

\bibitem{Akita:2020szl}
K.~Akita and M.~Yamaguchi, {\it {A precision calculation of relic neutrino
  decoupling}},  {\it JCAP} {\bf 08} (2020) 012,
  [\href{http://arxiv.org/abs/2005.07047}{{\tt arXiv:2005.07047}}].

\bibitem{Froustey:2020mcq}
J.~Froustey, C.~Pitrou, and M.~C. Volpe, {\it {Neutrino decoupling including
  flavour oscillations and primordial nucleosynthesis}},  {\it JCAP} {\bf 12}
  (2020) 015, [\href{http://arxiv.org/abs/2008.01074}{{\tt arXiv:2008.01074}}].

\bibitem{Bennett:2020zkv}
J.~J. Bennett, G.~Buldgen, P.~F. De~Salas, M.~Drewes, S.~Gariazzo, S.~Pastor,
  and Y.~Y.~Y. Wong, {\it {Towards a precision calculation of $N_{\rm eff}$ in
  the Standard Model II: Neutrino decoupling in the presence of flavour
  oscillations and finite-temperature QED}},  {\it JCAP} {\bf 04} (2021) 073,
  [\href{http://arxiv.org/abs/2012.02726}{{\tt arXiv:2012.02726}}].

\bibitem{Li:2021okx}
S.-P. Li, X.-Q. Li, X.-S. Yan, and Y.-D. Yang, {\it {Simple estimate of BBN
  sensitivity to light freeze-in dark matter}},  {\it Phys. Rev. D} {\bf 104}
  (2021), no.~11 115007, [\href{http://arxiv.org/abs/2106.07122}{{\tt
  arXiv:2106.07122}}].

\bibitem{Benson:2014qhw}
{\bf SPT-3G} Collaboration, B.~A. Benson et~al., {\it {SPT-3G: A
  Next-Generation Cosmic Microwave Background Polarization Experiment on the
  South Pole Telescope}},  {\it Proc. SPIE Int. Soc. Opt. Eng.} {\bf 9153}
  (2014) 91531P, [\href{http://arxiv.org/abs/1407.2973}{{\tt
  arXiv:1407.2973}}].

\end{thebibliography}\endgroup

\end{document}